\documentclass[twocolumn,pre]{revtex4}

\usepackage{dcolumn}
\usepackage{amsmath}

\usepackage{graphicx}

\begin{document}

\newcommand{\dd}{\mathrm{d}}
\newcommand{\Ord}{\mathrm{O}}
\newcommand{\e}{\mathrm{e}}
\newcommand{\half}{\mbox{$\frac12$}}
\newcommand{\set}[1]{\lbrace#1\rbrace}
\newcommand{\av}[1]{\langle#1\rangle}
\newcommand{\etal}{{\it{}et~al.}}
\newcommand{\defn}{\textit}
\newcommand{\Beta}{\mathrm{B}}

\newlength{\figurewidth}
\setlength{\figurewidth}{0.95\columnwidth}
\setlength{\parskip}{0pt}
\setlength{\tabcolsep}{6pt}
\setlength{\arraycolsep}{2pt}

\title{The first-mover advantage in scientific publication}
\author{M. E. J. Newman}
\affiliation{Department of Physics and Center for the Study of Complex
Systems,\\
University of Michigan, Ann Arbor, MI 48109}

\begin{abstract}
  Mathematical models of the scientific citation process predict a strong
  ``first-mover'' effect under which the first papers in a field will,
  essentially regardless of content, receive citations at a rate enormously
  higher than papers published later.  Moreover papers are expected to
  retain this advantage in perpetuity---they should receive more citations
  indefinitely, no matter how many other papers are published after them.
  We test this conjecture against data from a selection of fields and in
  several cases find a first-mover effect of a magnitude similar to that
  predicted by the theory.  Were we wearing our cynical hat today, we might
  say that the scientist who wants to become famous is better off---by a
  wide margin---writing a modest paper in next year's hottest field than an
  outstanding paper in this year's.  On the other hand, there are some
  papers, albeit only a small fraction, that buck the trend and attract
  significantly more citations than theory predicts despite having
  relatively late publication dates.  We suggest that papers of this kind,
  though they often receive comparatively few citations overall, are
  probably worthy of our attention.
\end{abstract}
\maketitle

In an influential paper published in 1965, the
physicist-turned-historian-of-science Derek de Solla Price presented one of
the first quantitative studies---perhaps \emph{the} first---of patterns of
citations between learned papers~\cite{Price65}.  His article, entitled
\textit{Networks of scientific papers,} depicted papers as the nodes of a
network of information, linked together by the citations between them.  He
noted a number of striking features of this network, chief among them the
skewed distribution of citation frequency under which most papers received
only a small number of citations but there was a ``long tail'' consisting
of a few papers cited very many times.  More specifically he showed that
the fraction $p_k$ of papers within his sample that were cited exactly~$k$
times diminished with increasing~$k$ according to a Pareto distribution or
power law $p_k\sim k^{-\alpha}$ with $\alpha$ a constant, a result that has
since been confirmed in other larger studies~\cite{Seglen92,Redner98}.  In
the jargon of modern network analysis, the citations between papers form a
\textit{scale-free network}~\cite{BA99b}.

In a follow-up paper published a few years later, Price~\cite{Price76}
proposed a remarkably simple explanation for this observation.  He
suggested that citation was subject to what he called a \textit{cumulative
  advantage} process~\cite{note1}, whereby papers that had been cited many
times in the past were more likely to be cited again, resulting in a
compound interest effect under which the best cited papers became ever
better cited, leaving their less popular counterparts behind.  Price
proposed a mathematical model of this process and solved it to show that
indeed it gives rise to a power-law distribution of citation frequencies.

Inspired, among other things, by the distribution of links between pages on
the world wide web, which also appears to follow a power law, a similar
process---now with the new name of \textit{preferential attachment}---was
independently proposed by Barab\'asi and Albert in 1999~\cite{BA99b} and
elaborated upon extensively by a number of
authors~\cite{AB02,DM02,Newman03d}.  As a result of this and subsequent
work, the mathematics of preferential attachment and the power-law
distributions it produces is now quite well understood.

The preferential attachment mechanism is qualitatively plausible in
citation networks---one can certainly imagine that papers cited many times
in the past are more likely to be read and cited again.  There is also good
empirical evidence in its favor~\cite{JNB03}, although deviations from its
predictions have been observed in some data
sets~\cite{BMG04,LJ05,Redner05}, especially those that span large periods
of time.  Overall, however, the preferential attachment mechanism is widely
considered to be a reasonable, if simplistic, explanation for the long tail
in citation networks.

What is less widely appreciated, perhaps, is that preferential attachment,
if correct, would also imply a variety of other distinctive features in
citation networks that could, at least in principle, be observed
empirically.  In particular, as we discuss here, it should produce a strong
\textit{first-mover advantage}, under which the first papers published on a
topic should enjoy far higher rates of citation, in perpetuity, than those
that come after them.  That preferential attachment should imply some kind
of first-mover advantage has been pointed out by Adamic and
Huberman~\cite{AH00a}, who looked for such effects in web data but found no
evidence that they existed.  In scientific citation, by contrast, it is
almost axiomatic that the first papers in a field are important and should
be highly cited~\cite{Redner05}, although it's unclear how large the effect
should be.  In this paper we calculate the size of the first-mover effect
within the preferential attachment model and compare the results with
citation data from a number of fields.  We find not only that scientific
citation shows a substantial first-mover effect, but that the size and
duration of the effect often agree closely with the theoretical
predictions.

\section*{The first-mover advantage}
In the model of citation proposed by Price~\cite{Price76}, it is assumed
that on average each paper published cites $m$ previous papers, which are
chosen in proportion to the number~$k$ of citations they already have plus
a positive constant~$r$.  The constant is necessary to ensure that papers
with no previous citations---which is all papers upon first
appearance---can still receive citations.  Price himself studied only the
case $r=1$, but the generalization to other cases is trivial.  The model
also includes the widely studied model of Barab\'asi and
Albert~\cite{BA99b} as a special case when $r=m$.

The model can be solved exactly for the citation distribution in the limit
of a large number of papers using a master-equation method introduced by
Simon~\cite{Simon55,Price76,KR01}.  The solution shows that the
fraction~$p_k$ of papers with exactly $k$ citations in the model network is
\begin{equation}
p_k = {\Beta(k+r,\alpha)\over\mathrm{B}(r,\alpha-1)},
\label{eq:pricepk}
\end{equation}
where
\begin{equation}
\Beta(a,b) = {\Gamma(a)\Gamma(b)\over\Gamma(a+b)}
\end{equation}
is the Euler beta function, $\Gamma(a)$~is the standard gamma function, and
\begin{equation}
\alpha = 2 + {r\over m}.
\label{eq:alpha}
\end{equation}
For large values of its first argument, the beta function has the
asymptotic form $\Beta(a,b)\sim a^{-b}$, and hence the tail of the citation
distribution has a power-law form $p_k\sim k^{-\alpha}$, just as in the
empirical data.

There is much more that can be done with the model, however, than just
calculating total numbers of citations.  In particular, one can calculate
the full distribution of citations a paper should receive as a function of
its date of publication.

Suppose $n$ papers in all have been published and let $p_k(i,n)$ be the
probability that the $i^\mathrm{th}$ paper has exactly $k$ citations.
Then, as shown for the case $m=1$ by Krapivsky and Redner~\cite{KR01} and
more generally by Dorogovtsev and Mendes~\cite{DM02}, $p_k(i,n)$~satisfies
the master equation
\begin{align}
(n+1) p_k(i,n+1)
  &= np_k(i,n) \nonumber\\
  & \hspace{-7em} {} + {m\over m+r} \bigl[ (k-1+r) p_{k-1}(i,n)
                                 - (k+r) p_k(i,n) \bigr].
\label{eq:agerate1}
\end{align}
The only exception to this equation is for the case~$k=0$, where instead we
get
\begin{equation}
(n+1) p_0(i,n+1)
  = np_0(i,n) + \delta_{in} - {mr\over m+r} p_0(i,n).
\label{eq:agerate0}
\end{equation}
Notice the Kronecker delta, which adds a single paper with $k=0$ if $i=n$,
but none otherwise.

In~\cite{KR01,DM02} these equations are solved using a continuous-time
method in which the discrete difference equation is approximated by a
differential equation.  Here we take a slightly different approach that
leads to the same conclusions but avoids the approximation.  Let us define
a ``time'' variable~$t$ such that the $i^\mathrm{th}$ paper published has
$t=i/n$.  (Technically $t$ need have nothing to do with actual time---it
measures only the sequence of publications---so the model can be used even
if the rate of appearance of publications is not constant over time, a
common situation.)  At the same time let us also change from $p_k(i,n)$ to
a density function $\pi_k(t,n)$ such that $\pi_k(t,n)\>\dd t$ is the
expected fraction of papers that have been cited~$k$ times and fall in the
interval from $t$ to $t+\dd t$.  Substituting for these quantities in
Eqs.~\eqref{eq:agerate1} and~\eqref{eq:agerate0}, taking the limit of
large~$n$, and introducing the shorthand notation $\pi_k(t)$ for the
limiting distribution~$\pi_k(t,\infty)$, we find that
\begin{equation}
(k+r) \pi_k(t) - (\alpha-1)\, t {\dd\pi_k\over\dd t} =
   (k-1+r) \pi_{k-1}(t),
\label{eq:agediff1}
\end{equation}
with the convention that $\pi_{-1}(t)=0$ for all~$t$, $\alpha$~defined as
in~\eqref{eq:alpha}, and boundary conditions $\pi_k(0)=\delta_{k,0}$.

It is straightforward to verify that these equations have the solution
\begin{equation}
\pi_k(t) = {\Gamma(k+r)\over\Gamma(k+1)\Gamma(r)}\,t^{r/(\alpha-1)}
               \bigl(1 - t^{1/(\alpha-1)}\bigr)^k,
\label{eq:pi1}
\end{equation}
which is equivalent to the solution given in~\cite{DM02}, and from this one
can calculate the average number~$\gamma(t)$ of citations a paper receives
as a function of its time of publication:
\begin{equation}
\gamma(t) = \sum_{k=0}^\infty k \pi_k(t)
          = r \bigl( t^{-1/(\alpha-1)} - 1 \bigr).
\label{eq:gamma}
\end{equation}

The first thing to notice about this expression is that it becomes
arbitrarily large as $t\to0$, meaning that, if the assumptions of the model
are correct, papers published early on should expect to receive far more
citations than those published later, even after allowing for the fact that
later papers have less time to accrue citations.  This is the first-mover
advantage.  In effect, as argued in~\cite{AH00a}, the model predicts not
only that there will be a long tail to the citation distribution, but that
the tail will be composed principally of the earliest papers published.

\begin{figure*}[t]
\begin{center}
\includegraphics[width=12cm]{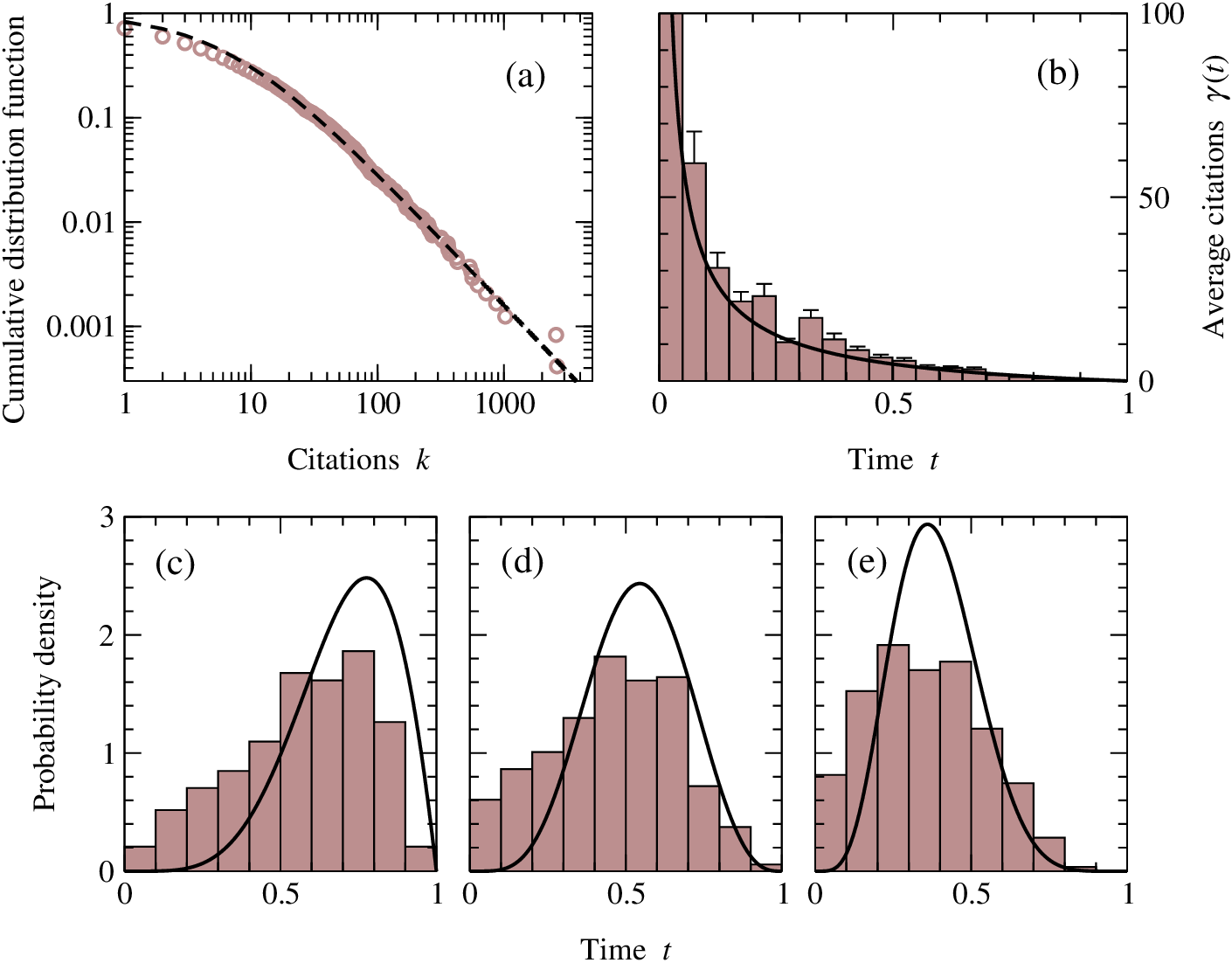}
\end{center}
\caption{Statistics of citations to papers about the theory of networks.
  Empirical measurements are in brown; theoretical predictions are in
  black.  (a)~Cumulative distribution of number of citations received by a
  paper.  The best fit to the theory is achieved for $\alpha=2.28$,
  $r=6.38$.  (b)~Mean number of citations received by papers as a function
  of time from beginning ($t=0$) to end ($t=1$) of the period covered.
  (c), (d)~and~(e): Probability that a paper with a given number of
  citations is published at time~$t$ for papers with (c)~1~or 2 citations,
  (d)~3~to 5 citations, and (e)~6~to 10 citations.}
\label{fig:networks}
\end{figure*}

The common-sense reasoning behind this observation is that papers published
early in a field receive citations essentially regardless of content
because they are the only game in town.  Authors feel the need to cite
\emph{something} and if there is only a small number of relevant
publications then inevitably those publications get cited.  This gives the
earliest publications a head start that is subsequently amplified by the
preferential attachment process and they continue to receive citations
indefinitely at a higher rate than later papers because they have more
citations to begin with.

There may be some truth to this.  It is not uncommon to hear a scientist
grumble about a paper that, in their opinion, receives citations only
because ``everybody cites that paper.''  More to the point, citing a paper
because it was the first in an area is in most cases entirely appropriate:
the first researcher to bring an issue to the attention of the scientific
community surely deserves credit for it, regardless of whether the other
details of their paper turn out later to be important.  On the other hand,
it is clear that the detailed contents of a paper usually do play a role in
determining the attention it receives~\cite{AL89,Redner05} and thus the
preferential attachment model cannot be a complete representation of the
citation process.  Nonetheless, as we now show, it turns out in some cases
to describe the statistical features of real citation networks with
surprising accuracy.

\section*{Comparison with citation data}
Testing the preferential attachment model against citation data poses some
challenges.  It is clear, for a start, that the model cannot be considered
seriously as a representation of the \emph{entire} citation network, the
network of all papers ever published, since the literature is divided into
many fields with most citations falling between papers in the same field.
In the best of worlds, therefore, the model could only reasonably be
considered to describe citations within a single field.  Moreover, given
that a large part of our interest here focuses on the qualities of the
earliest papers published, we need to find data that describe a field from
its earliest foundation and such data can be hard to come by.

Figure~\ref{fig:networks} shows one relatively clean example, a citation
network of papers on network theory---the topic of this very paper.  Within
the hard sciences this is a field of relatively recent provenance,
attracting an impressive volume of research since the late 1990s but almost
none before that.  (Its history in the social sciences is much longer, as
Price's work attests, but cross-citation between the areas is rare enough
as to be negligible---a lucky feature for our analysis, although an
unfortunate one for the progress of science.)

Our citation network for this example consists of five early and well cited
papers in the field~\cite{BA99b,WS98,AB02,DM02,Newman03d} along with all
papers that cite them, but excluding review articles, which tend to have
distinctively different citation patterns, and restricted to papers in
physics and related areas.  The resulting data set contains 2407 papers in
all spanning a ten-year period from June 1998 to June 2008.
Figure~\ref{fig:networks}a shows the complete distribution of numbers of
citations to these papers, along with the best fit to the
form~(\ref{eq:pricepk}).  As we can see, this two-parameter fit is
remarkably good, and allows us to extract accurate values for $r$ and
$\alpha$ from the data.  In Fig.~\ref{fig:networks}b, we show the average
number of citations received by papers as a function of time, where time is
measured, as in the model, in terms not of publication date but of
publication order.  We also show the predicted value of the same quantity
from Eq.~(\ref{eq:gamma}), and this is now, effectively, a zero-parameter
fit, since the two parameters appearing in the formula have already been
fixed.  Nonetheless, the agreement between data and theory is again good,
and in particular the data show a clear first-mover effect of a magnitude
and duration very similar to that predicted by the model.

As we mentioned above, the effect is sizable: the first 10\% of papers in
this example received an average of 101 citations each, while the second
10\%---published only a little later---received just~26.  The most recent
10\% of papers in the data set received a miserable 0.08 citations each,
meaning that most of them have never been cited at all.

\section*{Highly cited papers}
We can take our analysis further.
Figure~\ref{fig:networks}c--\ref{fig:networks}e shows a detailed comparison
of the actual distribution of citations at different times against the
theoretical predictions.  Again the fit is quite good, although there are
some interesting differences between data and model now visible.  In
particular, there are noticeably more papers published at early times in
each degree range than predicted by the theory and correspondingly fewer
around the peak value, meaning that not all papers in the early period are
benefiting from the first-mover advantage; we hope this is a positive sign
that citation rates are reflecting on paper content at least to some
extent.  More interestingly perhaps, there are also a scattering of papers
that are cited substantially more than expected.  These are few enough in
number that they have little visible effect on the figure, but the number
of citations they receive puts them well outside the expected range.

\begin{figure}
\begin{center}
\includegraphics[width=\columnwidth]{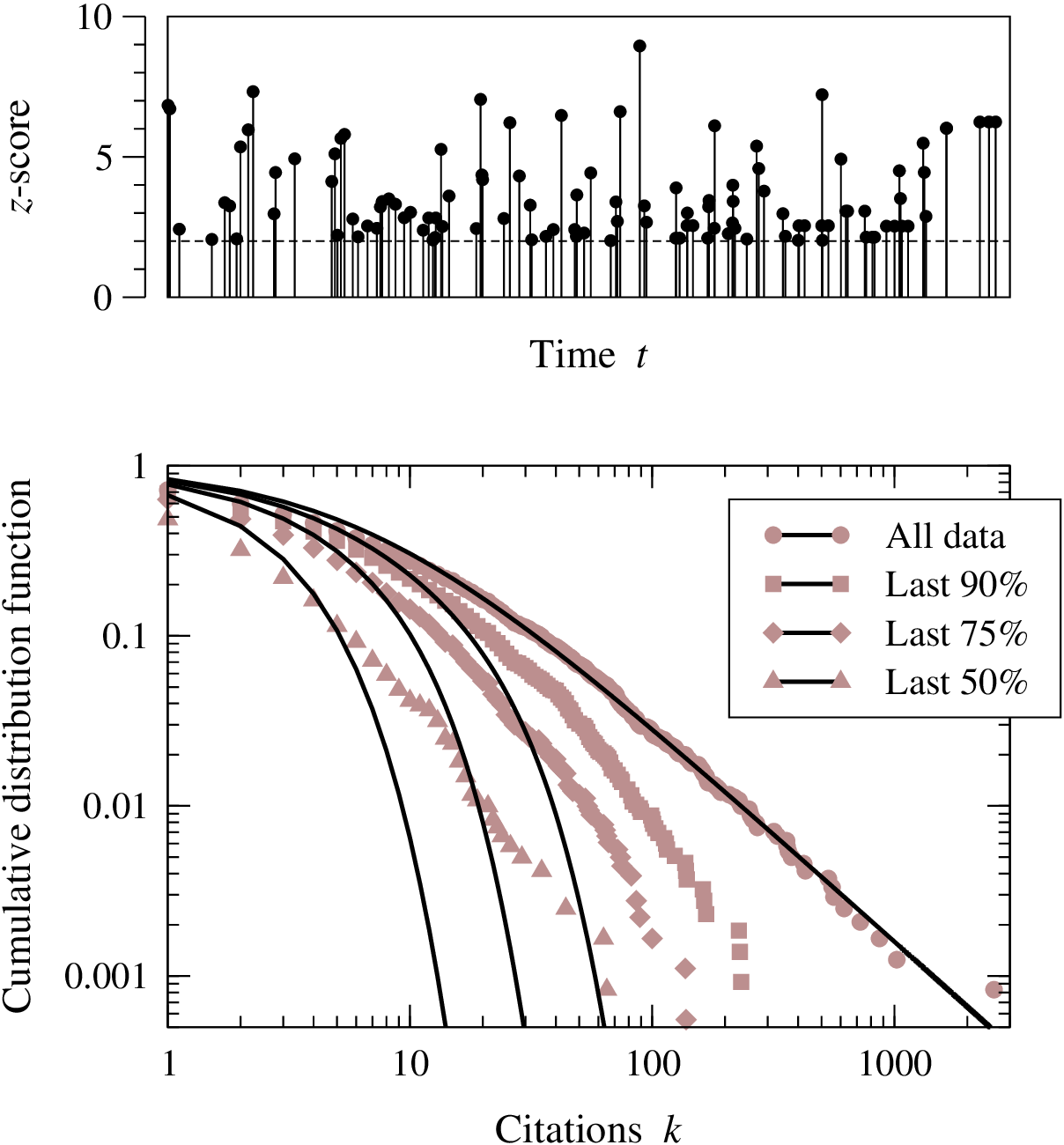}
\end{center}
\caption{Top: citations to papers in network theory measured in terms of
  number of standard deviations, or $z$-score, above the mean for papers
  published around the same date.  For clarity, only papers with $z$-score
  greater than~2 standard deviations (dotted line) are shown.  Bottom:
  cumulative distribution function for subsets of papers in the data set.
  The top curve represents the full data set, as in
  Fig.~\ref{fig:networks}a, while the other three represent the most recent
  90\%, 75\%, and 50\% of papers, respectively.  Points represent observed
  values and solid lines represent the theoretical prediction,
  Eq.~\eqref{eq:incomplete}.}
\label{fig:outliers}
\end{figure}

And this leads us to an interesting possibility.  It is common to assess
the importance of papers according to the number of citations they receive,
but the results presented here suggest that a large part of the variation
in citation numbers is a result of publication date rather than paper
content.  If, however, we measure a paper's citation count relative to the
average in its field \emph{for the given publication date}, then this
effect is factored out and---perhaps---the true stars of the citation
galaxy will emerge.  We have done this for our network theory data set,
calculating a $z$-score for each paper that measures the number of standard
deviations by which its citation count surpasses the appropriate
mean~\cite{note2}.  The ``best-cited'' papers by this measure turn out to
be roughly evenly distributed over the ten-year period covered by the data
set, as shown in Fig.~\ref{fig:outliers}, a sign that perhaps the technique
is indeed factoring out effects of timing.  Some early seminal works such
as Ref.~\cite{WS98}, which at 2623 citations is the most cited paper in the
data set, score highly ($7.1\sigma$ above the mean for its publication
date), but so do some later papers such as Ref.~\cite{BBPV04} ($6.5\sigma$
above the mean with 233 citations).  And Ref.~\cite{OHLN06} beats out both
of these at 7.2 standard deviations above the mean even though its
relatively recent 2006 publication date means it has received only 63
citations so far.  On the basis of these observations one might tentatively
predict that this paper (and others like it) will turn out to be
influential.

The appearance of well cited papers relatively late in the development of a
field is an encouraging sign that true citation patterns don't just
mindlessly follow the preferential attachment rule.  To quantify this
phenomenon further, we consider the overall distribution of numbers of
citations for papers in the latter part of our citation time series.
Suppose we are interested in papers published after some time~$t_0$.  Their
distribution within the preferential attachment model, which we'll
denote~$p_k(t_0)$, can be calculated by integrating Eq.~\eqref{eq:pi1}
thus:
\begin{align}
p_k(t_0) &= {1\over1-t_0}
       \int_{t_0}^1 \pi_k(t)\>\dd t \nonumber\\
    &\hspace{-2em} = {1\over(1-t_0)}\,{\Gamma(k+r)\over\Gamma(k+1)\Gamma(r)}
       \int_{t_0}^1 t^{r/(\alpha-1)}(1-t^{1/(\alpha-1)})^k \>\dd t.
\end{align}
With the substitution $u=1-t^{1/(\alpha-1)}$, the integral can be performed
and the complete result written in the form
\begin{equation}
p_k(t_0) = {I_{u_0}(k+1,\alpha+r-1)\over1-t_0}\,p_k(0),
\label{eq:incomplete}
\end{equation}
where
\begin{equation}
I_x(a,b) = {1\over\Beta(a,b)}\int_0^x u^{a-1} (1-u)^{b-1} \>\dd u
\label{eq:defsix}
\end{equation}
is the regularized incomplete beta function and $u_0=1-t_0^{1/(\alpha-1)}$.
Note that $p_k(0)$ is simply the distribution for the complete network
given in Eq.~\eqref{eq:pricepk}, so that the incomplete beta gives us the
factor by which the distribution for the later papers differs from the
complete distribution.

\begin{figure*}
\begin{center}
\includegraphics[width=12cm]{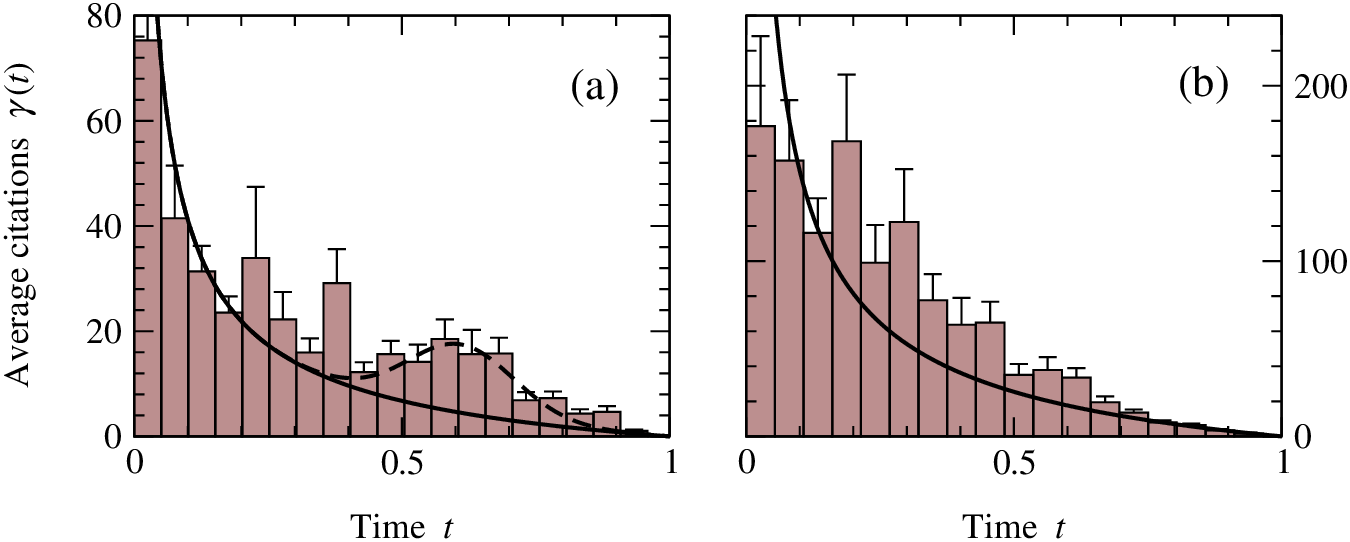}
\end{center}
\caption{Mean number of citations received as a function of time for papers
  on (a)~strangelets and strange matter and (b)~adult neural stem cells,
  along with theoretical predictions.  The foundational papers used to
  construct the two networks were \cite{Witten84,FJ84}
  and~\cite{RW92,RTW92}, respectively.}
\label{fig:other}
\end{figure*}

For large $a$ and fixed~$b$, as here, $I_x(a,b)$~has the asymptotic form
$I_x(a,b)\sim a^{b-1} x^a$~\cite{Gautschi67} and hence
\begin{equation}
I_{u_0}(k+1,\alpha+r-1) \sim k^{\alpha+r-2}
                             \bigl(1-t_0^{1/(\alpha-1)}\bigr)^k.
\end{equation}
Thus the citation distribution becomes exponentially truncated with a
typical scale
\begin{equation}
k_0 = - {1\over\ln \bigl(1-t_0^{1/(\alpha-1)}\bigr)},
\end{equation}
which for small values of $t_0$ is well approximated by
$k_0\simeq t_0^{-1/(\alpha-1)}$.

Put in simple terms, when we consider the citation distribution for just
the later papers in a field and exclude the earlier ones, we tend to throw
out the papers with the highest numbers of citations.  Thus one expects to
find a distribution in which few papers are highly cited.  When we look at
the data, however, we find poor quantitative agreement with the formulas
above in the tail of the distribution.  The lower panel in
Fig.~\ref{fig:outliers} shows the appropriate plot for our network science
example.  As the figure shows, the tail of the distribution is diminished
when we exclude the earliest papers in the data set, but by not as much as
the theory predicts.  Overall the tail still appears relatively long.  It
no longer follows the power-law form, but neither is it well described by
the predicted exponential.

Within the preferential attachment model it is almost impossible for later
papers in a network to get very many citations because there are only a
fixed number of citations to go around and most of them are going to the
earliest papers.  In Fig.~\ref{fig:outliers}, on the other hand, it is
clear that a substantial number of later papers are receiving large numbers
of citations, a hopeful sign that we as scientists do pay at least some
attention to good papers that come along later.

\section*{Other examples}
These analyses are for just one example field, which provides a
particularly clear instance of the first-mover effect.  The same methods
can be applied to other fields, though the results aren't often as clean.
There are a variety of issues that can complicate the analysis.  Some are
straightforward data problems: it may be difficult to restrict the set of
papers analyzed to those that truly fall in just one area, or to be sure
that you have captured all the relevant papers, or both.  It may even be
unclear when a field started at all (when did research on apple trees
begin?), or it may have started so long ago that modern concepts of
citation don't apply.

However, there are also some cases where differences between observation
and theory reveal behaviors of real scientific interest.  Two examples are
shown in Fig.~\ref{fig:other}.  Panel~(a) shows the curve of average
citation number for the subfield of particle physics concerned with
theories of ``strange matter''~\cite{Witten84,FJ84}---a topic that has at
the time of writing been receiving some attention with the start-up of the
Large Hadron Collider.  As the figure shows, there is again good general
agreement between observed citation counts and the model, and a strong
first-mover advantage similar in size to that predicted by the theory, an
interesting finding given that this data set spans an interval of 24
years---far longer than that of Fig.~\ref{fig:networks}.  However, there is
now also an additional ``bump'' in citation intensity in the latter half of
the time period, corresponding to papers published around 1999--2001 and
denoted roughly by the dotted line in the figure.  This bump, we assume, is
a result of true scientific developments in the field, though we leave the
experts in the area to suggest what developments those might be.

More substantial deviations from the theory arise when a branch of the
literature assumed to represent a new field turns out in fact to be merely
a subset of a larger, already-established field.  In this case, we would
not expect to see a first-mover effect at all.  The first papers published
in such a branch will be cited at a level typical of their position in the
middle of the larger subject area, and not as they would if they were the
only game in town.

We give an example of behavior of this type in Fig.~\ref{fig:other}b, which
shows citations to papers about adult neural stem cells.  The discovery
that neural stem cells exist not only during development but in adult
animals as well~\cite{RW92,RTW92} has resulted in a healthy quantity of
subsequent research, but it has not, at least according to our analysis,
created a ``new field.''  As Fig.~\ref{fig:other}b shows, the fit between
the observed citation record and the theory is poor in this case and in
particular the data show no discernible first-mover effect.  Earlier papers
in the data set do have more citations, but citation numbers appear to
increase only linearly with paper age, suggesting that on average papers
are being cited at roughly the same rate regardless of when they were
published.  A qualitative inspection of the data set indicates that in fact
the adult neural stem cell literature forms just a part of a larger
community of citation on neural stem cells in general and hence we indeed
expect no first-mover effect in this case.

Analyses of the type described here could thus, in principle, provide an
independent test of claims, frequent in some areas, that a particular
publication or discovery has created a new field of science.  We repeat our
caution, however, that the data are not always as clean as we would like
and it is not always possible to make a firm statement one way or the
other.

\section*{Conclusions}
In conclusion, the strong first-mover advantage predicted by theories of
the scientific citation process seems to be quantitatively substantiated by
empirical citation data, at least in some areas.  The cynical observer
would, it appears, have some justification in concluding that if you want
to be well cited you are better off writing the first paper in such an area
than writing the best.  Other areas, by contrast, show no first-mover
effect, which may be an indication that those areas do not constitute
self-contained research fields as assumed by the theory.  And even in cases
where the first-mover effect is strong, a small number of later papers do
seem buck the trend and attract significant attention in defiance of
predictions.  We tentatively suggest that the reader looking for true
breakthroughs could do worse than keep an eye out for papers such as these.

\begin{acknowledgments}
  The author thanks Carrie Ferrario, Loet Leydesdorff, Robert Messing,
  Aaron Pierce, and Sidney Redner for useful suggestions.  This work was
  funded in part by the National Science Foundation under grant
  DMS--0405348 and by the James~S. McDonnell Foundation.
\end{acknowledgments}


\begin{thebibliography}{10}
\bibitem{note1}
A related phenomenon, the ``Matthew effect,'' was previously discussed by
  Merton~\cite{Merton68}, who observed that the simultaneous discovery of
  an important result by two different scientists is more likely to be
  credited to the more famous of the two, so that fame begets more fame.
\bibitem{note2}
Mean and standard deviation are calculated as averages over papers in a
  Gaussian-weighted window of width 100 papers around the paper of interest.
\end{thebibliography}
\end{document}